\documentclass[twocolumn,aps]{revtex4}
\usepackage{hyperref}
\usepackage{graphicx}
\hypersetup{backref=true}
 
\usepackage{amsmath}
\usepackage{amssymb}

\def\F{{\cal F}}

\def\K{{\cal K}}

\def\N{{\cal N}}

\font\tenscr=rsfs10 scaled1100
\font\sevenscr=rsfs7 
\font\fivescr=rsfs5 
\skewchar\tenscr='177 
\skewchar\sevenscr='177
\skewchar\fivescr='177 
\newfam\scrfam 
\textfont\scrfam=\tenscr
\scriptfont\scrfam=\sevenscr 
\scriptscriptfont\scrfam=\fivescr

\def\linebreak{\hfill\break}

\def\bra<#1|{\langle #1\rvert}
\def\ket|#1>{\lvert#1 \rangle}
\def\braket<#1|#2>{\langle #1|#2 \rangle}

\def\therefore{\mbox{\setbox0=\hbox{X}\hbox{$\ldotp$}\raise0.7\ht0\hbox{$\ldotp$}\hbox{$\ldotp$}} \quad }
\def\because{\mbox{\setbox0=\hbox{X}\raise0.7\ht0\hbox{$\ldotp$}\hbox{$\ldotp$}\raise0.7\ht0\hbox{$\ldotp$}}\kern0pt }

\def\SO{{\rm SO}}

\def\ISO{{\rm ISO}}

\def\Frac(#1/#2){\left(\frac{#1}{#2}\right)}

\def\insbra#1{\left[ #1 \right]}

\def\Eq#1{\begin{equation} #1 \end{equation}}

\def\Eqr#1{\begin{eqnarray} #1 \end{eqnarray}}

\def\Eqrsubl#1#2{\begin{subequations}\label{#1}\Eqr{#2}\end{subequations}}

\def\Bitm{\begin{itemize}}
\def\Eitm{\end{itemize}}
\def\Blist#1#2{\begin{list}{#1}{\parsep=0pt \itemsep=0pt%
  \listparindent=0pt #2}}
\def\Elist{\end{list}}
\long\def\ignore#1#2{\def\ignoreflag{#1}\long\def\tmptext{#2}
  \ifnum\ignoreflag>1 #2 \fi}

\def\THB{{\mathbb T}}
\def\VHB{{\mathbb V}}
\def\SHB{{\mathbb S}}
\def\GHB{{\mathbb Y}}

\begin{document}

\title{
Uniqueness and Stability of Higher-Dimensional Black Holes%
\footnote{To appear in the proceedings of the 6th APCTP International Conference on Gravitation and Astrophysics.}}
\author{Hideo Kodama}%
 \email{kodama@yukawa.kyoto-u.ac.jp}
\affiliation{
Yukawa Institute for Theoretical Physics, Kyoto University\\
Kyoto 606-8502, Japan
}

\date{\today}

\begin{abstract}
The present status of the uniqueness and stability issue of black 
holes in four and higher dimensions is overviewed with focus on the 
perturbative analysis of this issue for static black holes in higher 
dimensions as well as those in four dimensions with cosmological 
constant.
\end{abstract}

\maketitle

\section{Introduction}

Does a gravitational collapse always lead to formation of a black 
hole? What kind of black holes can exist in nature?  Can realistic 
black holes be well described by stationary solutions? These are 
among the most important problems in general relativity and have 
been studied for long time.  Although we do not have a definite 
answer to the first question related to the cosmic censorship yet, 
we have rather well established answers to the second and third 
questions, which are summarised as the uniqueness and stability of 
asymptotically flat stationary regular black hole solutions. 

In local astrophysical problems in the present universe, the 
assumption of the asymptotic flatness will be a good approximation. 
However, the success of the inflationary universe model implies that 
the cosmological constant cannot be neglected even locally in the 
early universe. Further, recent developments in unifying theories 
suggest the possibility that the universe has dimensions higher than 
four on microscopic scales%
\cite{Antoniadis.I&&1998,Randall.L&Sundrum1999} 
and that mini black holes might be produced in elementary particle 
processes in colliders as well as in high energy cosmic shower 
events. Therefore, it is a quite important problem whether the 
uniqueness and stability of regular black hole solutions also hold 
in the non-vanishing cosmological constant case and/or in higher 
dimensions.

In this article, we briefly overview the present status of the 
investigations of this problem and discuss what kind of new 
information can be obtained by the linear perturbation theory. 

\section{Four-Dimensional Black Holes}

\subsection{Uniqueness of black holes}

\subsubsection{Asymptotically flat Einstein-Maxwell system}

As mentioned in Introduction, we now have an almost complete list of 
regular stationary black hole solutions for the asymptotically flat 
Einstein-Maxwell system (for historical references, see the reviews 
Refs. \cite{Heusler.M1996B,Chrusciel.P1996}). 

What has played the most important role in this classification 
problem is the rigidity theorem, which asserts that for the 
Einstein-Maxwell system, an asymptotically flat stationary regular 
analytic black hole solution whose horizon is an analytic 
submanifold is axisymmetric if it is rotating, i.e., the Killing 
vector that represents the time translation at infinity is not 
parallel to the null generators on the 
horizon\cite{Chrusciel.P1996}. The first full proof of this theorem 
was given in the textbook by Hawking and 
Ellis\cite{Hawking.S&Ellis1973B} under the assumption that the 
horizon is connected and non-degenerate, i.e., the surface gravity of the horizon does not vanish. Incompleteness of their 
proof concerning the claims that the domain of outer communication 
is simply connected and the black hole surface has $S^2$ topology 
was later removed by Chru\'sciel and 
Wald\cite{Chrusciel.P&Wald1994}. Further, the theorem was extended 
by Chru\'sciel\cite{Chrusciel.P1996} to include the case in which 
the horizon has multiple components and/or is degenerate. At about 
the same time, it was proved by R\'acz and 
Wald\cite{Racz.I&Wald1996} that a non-rotating regular black hole 
solution is static if the horizon is non-degenerate. 

Thus, the classification problem of asymptotically flat stationary 
(regular analytic) black hole solutions was reduced to that of 
static black holes and axisymmetric black holes. For neutral static 
black holes and charged static black holes with non-degenerate 
horizon, this classification problem is completely solved: the 
Schwarzschild and Reissner-Nordstr\"om solution are the only regular 
black hole solutions, as is well-known. In contrast, in the charged 
degenerate case, there exists in addition the family of the 
Majumdar-Papapetrou multiple black hole solutions, which was 
recently shown to exhaust all regular solutions with multiple black 
holes if the charges of all black holes have the same 
sign\cite{Chrusciel.P1999}. 

Next, in the rotating case, the Kerr-Newman solution is the only 
asymptotically flat regular solution, if the horizon is connected. 
Concerning multiple black hole solutions, Weinstein classified 
solutions that are regular except on the symmetry axis and found 
that for any positive integer $N$, there exist families of $N$ 
black hole solutions, with parallel spins aligned on the symmetry 
axis, which has $3N-1$ parameters for the vacuum 
case\cite{Weinstein.G1994} and $4N-1$ parameters for the 
electro-vacuum case\cite{Weinstein.G1996a}. At present, it is not 
known whether these families contains a completely regular solution 
without a strut singularity.

\subsubsection{Other asymptotically flat systems}

Uniqueness theorems are also proved for some asymptotically flat 
systems with other types of matter. For example, the only regular 
static black hole  solutions with non-degenerate horizon are given 
by the Gibbons-Maeda solutions\cite{Gibbons.G&Maeda1988} for the 
Einstein-Maxwell-Dilaton system%
\cite{Masood-ul-Alam.A1993,Mars.M&Simon2003,%
Gibbons.G&Ida&Shiromizu2002b,Rogatko.M1999,Rogatko.M2002}, 
and by the Reissner-Nordstr\"om solution for the 
Einstein-Maxwell-Dirac system. Further, it is also shown that the 
only regular stationary black hole solutions with non-degenerate 
horizon is the Kerr solution when the matter is described by a 
harmonic scalar\cite{Heusler.M1995}. Note that in this result, the 
rigidity theorem is assumed for the rotating case.

In the meantime, it is now known that the uniqueness theorem is not 
universal and does not hold for some systems. For example, for the 
Einstein-Yang-Mills system, it is shown that there exist a family of 
spherically symmetric regular soliton solutions and regular black 
hole solutions different from Schwarzschild solutions%
\cite{Bartnik.R&McKinnon1988,Volkov.M&Galtsov1990,%
Kunzle.H&Masood-ul-Alam1990}. Although these additional solutions are 
shown to be unstable%
\cite{Straumann.N&Zhou1990,Bizon.P1991,Zhou.Z&Straumann1991}, the 
Einstein-Skyrme system allows of a spherically symmetric regular 
black hole solution with a Skyrme 
hair\cite{Droz.S&Heusler&Straumann1991} that is stable%
\cite{Heusler.M&Droz&Straumann1991,Heusler.M&Straumann&Zhou1993}. 

\subsubsection{Asymptotically dS and AdS systems}

The investigation of the classification of black hole solutions had 
been mainly concerned with the asymptotically flat case for a long 
time. This was because basic tools used to prove the uniqueness 
theorems for the asymptotically flat case do not work for systems 
with non-vanishing cosmological constant. 

Recently, however, Anderson has worked out a new approach which 
leads to a new type of rigidity theorem and enables us to study the 
classification problem of static spacetimes with negative 
cosmological constant%
\cite{Anderson.M2001A,Anderson.M2001Aa}. Utilising this approach, 
Anderson, Chru\'sciel and Delay showed that an asymptotically 
anti-de Sitter and static vacuum solution containing one 
non-degenerate regular black hole is unique and given by the 
Schwarzschild-AdS black hole solution if it has a $C^2$ conformal 
spatial completion\cite{Anderson.M&Chrusciel&Delay2002}. 
Unfortunately, their method cannot be applied to the rotating case 
or to the case of positive cosmological constant. As far as the 
author knows, no uniqueness theorem with sufficient generality has 
been proved in the case of positive cosmological constant. Later, we 
will show that in a perturbative sense, the uniqueness of black hole 
solutions can be proved even for this positive cosmological constant 
case.

\subsection{Stability of black holes}

The uniqueness of black holes does not necessarily imply their 
stability. In particular, if the cosmic censorship hypothesis (CCH) 
does not hold, a regular stationary black hole solution will be 
unstable against a generic perturbation, because formation of naked 
singularity must be generic\cite{Wald.R1997A}. 

In reality, many of the standard black hole solutions are known to 
be perturbatively stable, supporting CCH. For example, both 
Schwarzschild black holes and non-degenerate Reissner-Nordstr\"om 
black holes are proved to be stable%
\cite{Vishveshwara.C1970,Price.R1972,Wald.R1979,Wald.R1980,%
Chandrasekhar.S1983B}. Recently, the Kottler black holes, i.e., the 
spherically symmetric black holes in spacetimes with non-vanishing 
cosmological constant, were also shown to be stable by the author 
and 
Ishibashi\cite{Kodama.H&Ishibashi2004}. 

In contrast to static black holes, the stability of rotating black 
holes are established only for the Kerr black 
holes\cite{Whiting.B1989}. We do not discuss the stability of these 
rotating black holes in this article.

In the proofs of the stability of these black holes, a key role was 
played by the fact that the basic equations for perturbations can be 
reduced to a single second-order ODE (ordinary differential 
equation) of the Schr\"odinger type, which is often called a master 
equation. Now, we explain how such a master equation is derived and 
how the stability is proved in terms of it.

\subsubsection{Tensorial decomposition of perturbations}

In spacetimes of dimension $n+2$, the linearisation of the Einstein 
equations yields the following equations for the metric perturbation 
$h_{\mu\nu}=\delta g_{\mu\nu}$:
\Eqr{
&& (\triangle_L h)_{\mu\nu}-\nabla_\mu\nabla_\nu h
   +2\nabla_{(\mu}\nabla^\alpha h_{\nu)\alpha}
   \notag\\
&&\qquad
  +(-\nabla^\alpha\nabla^\beta h_{\alpha\beta}
   +\triangle h +R^{\alpha\beta}h_{\alpha\beta})g_{\mu\nu}
   \notag\\
&&\qquad 
   +(2\Lambda-R)h_{\mu\nu} =2\kappa^2\delta T_{\mu\nu},
}
where $\triangle_L$ is the Lichinerowicz operator defined by
\Eq{
(\triangle_L h)_{\mu\nu}:=-\nabla\cdot\nabla h_{\mu\nu}
  +2R_{(\mu}^\alpha h_{\nu)\alpha}
  -2R_{\mu\alpha\nu\beta}h^{\alpha\beta}.
}

In order to analyse the behaviour of perturbations on the basis of 
these equations, the following two technical problems have to be 
resolved. Firstly, the perturbation variables $h_{\mu\nu}$ contain 
unphysical gauge degrees of freedom that should be eliminated. 
Second, these perturbation equations are a coupled system of a large 
number of equations and are quite hard to solve directly. 

In the case of our interest in which the unperturbed background 
spacetime is spherically symmetric, these problems can be resolved 
with the help of the tensorial decomposition and the gauge-invariant 
formulation%
\cite{Bardeen.J1980,Kodama.H&Sasaki1984,Kodama.H&Ishibashi&Seto2000}. 
First, the metric of a spherically symmetric black hole solution to 
the vacuum or electro-vacuum system in $n+2$ dimensions is written 
as 
\Eq{
ds^2= g_{ab}(y)dy^ady^b + r^2(y) d\Omega^2,
}
where $g_{ab}dy^a dy^b$ is the metric of a two-dimensional orbit 
space $\N^2$, which can be written as $-f(r)dt^2+f(r)^{-1}dr^2$, and 
$d\Omega^2= \gamma_{ij}(z)dz^idz^j$ is the metric of the unit sphere 
$S^n$. On this background, the basic perturbation variables 
$h_{\mu\nu}$ can be grouped into three sets, $h_{ab}$, $h_{ai}$ and 
$h_{ij}$, according to their tensorial transformation behaviour on 
$S^n$. Among these, $h_{ai}$ can be further decomposed into the 
gradient part $h_a$ and the divergence-free part $h_{ai}^{(1)}$ as
\Eq{
h_{ai}=\hat D_i h_a + h_{ai}^{(1)};\quad
\hat D^i h_{ai}^{(1)}=0,
}
where $\hat D_i$ is the covariant derivative with respective to the 
metric $\gamma_{ij}$ on $S^n$. Similarly, $h_{ij}$ can be decomposed 
into the trace part and 
trace-free part as
\Eq{
h_{ij}=h_L \gamma_{ij} + h_{Tij};\quad
h_{Tj}^j=0,
}
and the trace-free part $h_{Tij}$ can be further decomposed into the 
scalar part $h_T^{(0)}$, the divergence-free vector part 
$h_{Ti}^{(1)}$ and the divergence-free and trace-free part 
$h_{Tij}^{(2)}$ as
\Eqr{
&& h_{Tij}=\left(  \hat D_i\hat D_j
      -\frac{1}{n}\gamma_{ij}\hat\triangle\right)h^{(0)}_{T}
   +2\hat D_{(i} h^{(1)}_{T j}) +h^{(2)}_{T ij};\notag\\ 
&& \hat D^j h^{(1)}_{T j}=0,\ 
   \hat D^j h^{(2)}_{T ij}=0. 
}

By this decomposition, the linearised Einstein equations are 
decomposed into three decoupled sets of equations, each of which 
contains only variables belonging to one of the three sets of 
variables, the scalar-type variables $(h_{ab}, h_a, h_L, 
h_T^{(0)})$, vector-type variables $(h^{(1)}_{ai}, h^{(1)}_{Ti})$, 
and tensor-type variables $(h^{(2)}_{Tij})$. Further, since the 
covariant derivative $\hat D_i$ always appears in the combination 
$\hat\triangle := \hat D_i \hat D^i$ in the perturbation equations 
due to spherical symmetry, the expansion coefficients of these 
variables in terms of harmonic tensors on $S^n$, which are denoted 
as $(f_{ab},f_a^{(0)},H_L,H_T^{(0)})$, $(f_a^{(1)},H_T^{(1)})$ and 
$H_T^{(2)}$ for the scalar-type, vector-type and tensor-type 
variables, respectively, are mutually coupled among only those 
corresponding to the same harmonic tensor (see 
Ref.\cite{Kodama.H&Ishibashi2004,Gibbons.G&Hartnoll2002} for details 
of this harmonic expansion). Thus, the perturbation equations can be 
reduced to sets of equations on the two-dimensional orbit space 
$\N^2$ with small number of entries independent of the spacetime 
dimension. In addition, after this harmonic expansion, we can easily 
construct a basis for gauge-invariant variables by taking 
appropriate linear combinations of these two-dimensional variables 
and their derivatives with respect to the coordinates $y^a$ of 
$\N^2$, and the linearised Einstein equations can be written as 
differential equations for these gauge-invariant 
variables\cite{Kodama.H&Ishibashi&Seto2000}. Thus, we can 
essentially resolve the two technical problems mentioned at the 
beginning of this subsection. 

\subsubsection{Schwarzschild black hole}

By applying the above arguments to four-dimensional spherically 
symmetric black holes, we obtain two decoupled sets of equations for 
the scalar-type and vector-type perturbations, which are often 
called the even modes (or the polar perturbation) and the odd modes ( 
or the axial perturbation), respectively. The tensor-type 
perturbation does not exist because there is no trace-free and 
divergence-free symmetric tensor on $S^2$. Since the background is 
static, by the Fourier transformation with respect to the time 
coordinate $t$, 
\Eqr{
&& h_{ab}=f_{ab}(r)e^{-i\omega t}\GHB(\theta,\phi),\notag\\ 
&& h^{(1)}_{ia}=f_a(r)e^{-i\omega t}\GHB_i(\theta,\phi),\cdots,
}
these equations can be further reduced to sets of ODEs for functions 
of $r$. 

For the vector-type perturbation in the Schwarzschild background 
with $f(r)=1-2M/r$, the corresponding set of equations can be easily 
reduced to a single second-order ODE of the Shcr\"odinger type,
\Eq{
-\frac{d^2 \Phi}{dr_*^2}+ V(r)\Phi = \omega^2 \Phi;\quad
r_*= \int^r \frac{dr}{f(r)},
\label{MasterEq}
}
as first shown by Regge and Wheeler\cite{Regge.T&Wheeler1957}. It 
was shown by Zerilli later that a similar reduction is possible for 
the scalar-type perturbation of the Schwarzschild black hole as 
well\cite{Zerilli.F1970}. Although these authors derived these 
master equations by the gauge-fixing method, the master variable 
$\Phi$ can be  written in terms of gauge-invariant 
variables\cite{Kodama.H&Ishibashi2003}. Thus, the linear stability 
of the Schwarzschild black hole is determined by the behaviour of the 
effective potential $V(r)$ for each mode.

The explicit expressions for the effective potentials are given by
\Eq{
V_V=\frac{f}{r^2}\left( m+2-\frac{6M}{r} \right),
}
for the vector perturbation and 
\Eq{
V_S=\frac{f}{r^2H^2}\insbra{ m^2(m+2)+ \frac{6m^2M}{r}
      +\frac{36mM^2}{r^2}+\frac{72M^3}{r^3} },
}
for the scalar perturbation. Here, $m$ is written in terms of the 
eigenvalue of the harmonic scalar $\triangle \GHB=-l(l+1)\GHB$ as 
$m=(l-1)(l+2)$ ($l=2,3,\cdots$), and $H=m+6M/r$.

It is easy to see that the effective potentials $V_V$ and $V_S$ are 
positive outside the horizon $(r>2M)$. Because $r_*$ runs in the 
range $(-\infty,+\infty)$, this implies that the master equation for 
$\Phi$ has no regular and normalizable solution for $\omega^2<0$. 
Hence, the 4D Schwarzschild black hole is perturbatively 
stable\cite{Vishveshwara.C1970}.

\subsubsection{Reissner-Nordst\"om black hole}

For charged static black holes, metric perturbations couple  
perturbations of electromagnetic fields. However, by taking 
appropriate combinations of perturbation variables, the perturbation 
equations for each type can be reduced to two decoupled ODEs as
\Eqr{
-\frac{d^2 \Phi_\pm}{dr_*^2}+ V_\pm(r)\Phi_\pm = \omega^2 
\Phi_\pm,
\label{MasterEq:RN}
}
where $\Phi_+$ and $\Phi_-$ represent the electromagnetic and 
gravitational modes, respectively%
\cite{Moncrief.V1974,Zerilli.F1974,Chandrasekhar.S1983B}. 

The effective potentials $V_\pm$ can be shown to be  positive 
outside the horizon. Hence, the Reissner-Nordst\"om black hole is 
perturbatively stable outside the (outer Killing) horizon, although 
it is unstable even against spherically symmetric perturbations 
inside the outer horizon (mass inflation\cite{Poisson.E&Israel1990}).

\subsubsection{4D Kottler black holes}

Recently, Cardoso and Lemos found that the perturbation equations 
for 4D Schwarzschild-dS and -AdS black holes can be also reduced to 
a single master equation of the Schr\"odinger 
type\cite{Cardoso.V&Lemos2001}(see also 
Ref.\cite{Kodama.H&Ishibashi2003}). 

First, the effective potential for the scalar perturbation is given by
\Eqr{
&V_S=
 &\frac{f(r)}{r^2H^2}\left[ 18x^2f(r)+27x^3
         +9(2+5l_2+l_2^2)x^2\right.\notag\\
&&  \left. \qquad +(l_2+1)^2(l_2+4)^2(l_2^2+5l_2+9) \right],
}
where $l_2=l-2\ge0$, $x=2M/r$ and 
\Eq{
f(r)=1-\lambda r^2-\frac{2M}{r}.
}
Clearly, $V_S$ is positive definite outside the horizon ($f(r)>0$). 
Hence, there exists no normalizable mode with $\omega^2<0$.

Next, the effective potential for the vector perturbation is given by
\Eq{
V_V=\frac{f(r)}{r^2}\left[ 3f(r)+3+3\lambda r^2+l_2(l_2+5) \right].
}
This effective potential is positive definite outside the horizon 
for $\lambda\ge0$. However, for $\lambda<0$ it becomes negative in a 
neighbourhood of the horizon if $|\lambda|$ is large. Nevertheless, 
we can show that there exists no normalizable $\omega^2<0$ mode for 
vector perturbations as well. One way to show this is to utilise the 
scalar-vector correspondence explained next.

\subsubsection{The scalar-vector correspondence}

Solutions $\Phi_1$ and $\Phi_2$ of the same frequency $\omega$ to two ODEs of the Schr\"odinger equation type are related by
\Eqr{
\Phi_2(r)=p(r)\Phi_1(r)+ q(r)\partial_r \Phi_1(r),
} 
where $p(r)$ and $q(r)$ are functions of $r$ that do not depend on 
$\omega$, if and only if the effective potentials $V_1(r)$ and 
$V_2(r)$ for the two equations are expressed in terms of a single 
function $F(r)$ as 
\Eq{
V_1, V_2=\pm f\frac{dF}{dr}+F^2+cF,
}
for some constant $c$. Such functions exist for the Regge-Wheeler 
and Zerilli potentials%
\cite{Chandrasekhar.S&Detweiler1975,Chandrasekhar.S1983B}.

This scalar-vector correspondence also holds for the 
Schwarzschild-dS and -AdS black holes\cite{Cardoso.V&Lemos2001}. 
Hence, we can conclude that there exists no normalizable vector 
perturbation with $\omega^2<0$ from the above result for scalar 
perturbations.

\subsubsection{Perturbative uniqueness of 4D black holes} 
\label{PerturbativeUniqueness:4D}

We can obtain some information about the uniqueness problem of black 
holes by the perturbation analysis. This is because there must exist 
a bounded regular solution with $\omega=0$ to the perturbation 
equations if there were a continuous family of stationary regular 
black hole solutions that contains a spherically symmetric solution. 

For example, by a method called {\em $S$-deformation}, which is 
explained later, we can show that there exists no bounded regular 
solution with $\omega^2=0$ outside the horizon for neutral as well 
as charged spherically symmetric static black holes, except for  
stationary vector perturbations with $l=1$. These stationary 
solutions can be parametrised by three parameters that can be 
interpreted as the angular momentum of perturbations.  Hence, we can 
conclude that 4D asymptotically dS or AdS static regular black hole 
solutions are unique in a perturbative sense under the conditions 
that the spacetime is static and contains a single black hole. This 
argument also suggests that the extensions of the Kerr solution to 
the non-vanishing cosmological constant case, which are described by 
the neutral Carter solutions with vanishing NUT 
parameter\cite{Carter.B1968a}, are the only asymptotically dS or AdS 
stationary regular solutions with a connected horizon, in a vicinity 
of their static and spherically symmetric limit.

\section{Higher-Dimensional Black Holes}

\subsection{Generalised static solution}

It is quite easy to find the higher-dimensional counter-parts of 4D 
spherically symmetric black holes, such as the 
Tangherlini-Schwarzschild solution\cite{Tangherlini.F1963}. In 
reality, it is also possible to find slightly more general family of 
solutions for the 
Einstein-Maxwell system with cosmological constant $\Lambda$ by 
requiring that the spacetime metric and the electromagnetic tensor 
$\F$ can be expressed in the form
\Eqrsubl{GeneralisedStaticSol}{
&& ds^2= g_{ab}(y)dy^ady^b + r^2(y) d\sigma_n^2, 
\label{GeneralisedStaticMetric}\\
&& \F=\frac{1}{2} E_0 \epsilon_{ab} dy^a\wedge dy^b,
\label{GeneralisedStaticEM}
}
where $g_{ab}dy^a dy^b$ is a metric of a two-dimensional space 
$\N^2$, and $\sigma_n^2=\gamma_{ij}(z)dz^i dz^j$ is a metric of an 
Einstein space $\K^n$ satisfying
\Eq{
\hat R_{ij}=(n-1)K \gamma_{ij},
}
with $K=0,\pm1$. The Einstein equations and Maxwell equations 
determine the two-dimensional metric $g_{ab}$ and the electric field 
$E_0$ as%
\cite{Birmingham.D1999,Kodama.H&Ishibashi2004}
\Eq{
g_{ab}dy^a dy^b= -f(r)dt^2+\frac{dr^2}{f(r)},\quad
E_0=\frac{q}{r^n},
\label{GeneralisedStaticSol:f&E}
}
where
\Eqr{
&& f(r)=K -\lambda r^2 -\frac{2M}{r^{n-1}}+\frac{Q^2}{r^{2n-2}},\\
&& \lambda=\frac{2\Lambda}{n(n+1)},\quad
   Q^2=\frac{\kappa^2q^2}{n(n-1)}.
}

In the special case in which $\K^n$ is the unit sphere $S^n$, the 
spacetime becomes static and spherically symmetric. In particular, 
for $K=1,\Lambda=0$ and $Q=0$, this solution coincides with the 
Tangherlini-Schwarzschild solution, and for $K=1$ and $\Lambda=0$, 
it gives an higher-dimensional counter-part of the 
Reissner-Nordstr\"om solution. More generally, when $\K^n$ is a 
constant curvature space, the solution is invariant under 
$\SO(n+1)$, $\ISO(n)$ and $\SO(n,1)$ for $K=1,0,-1$, respectively.

\begin{figure}[t]
\begin{minipage}{4cm}
\centerline{\includegraphics[width=4cm]{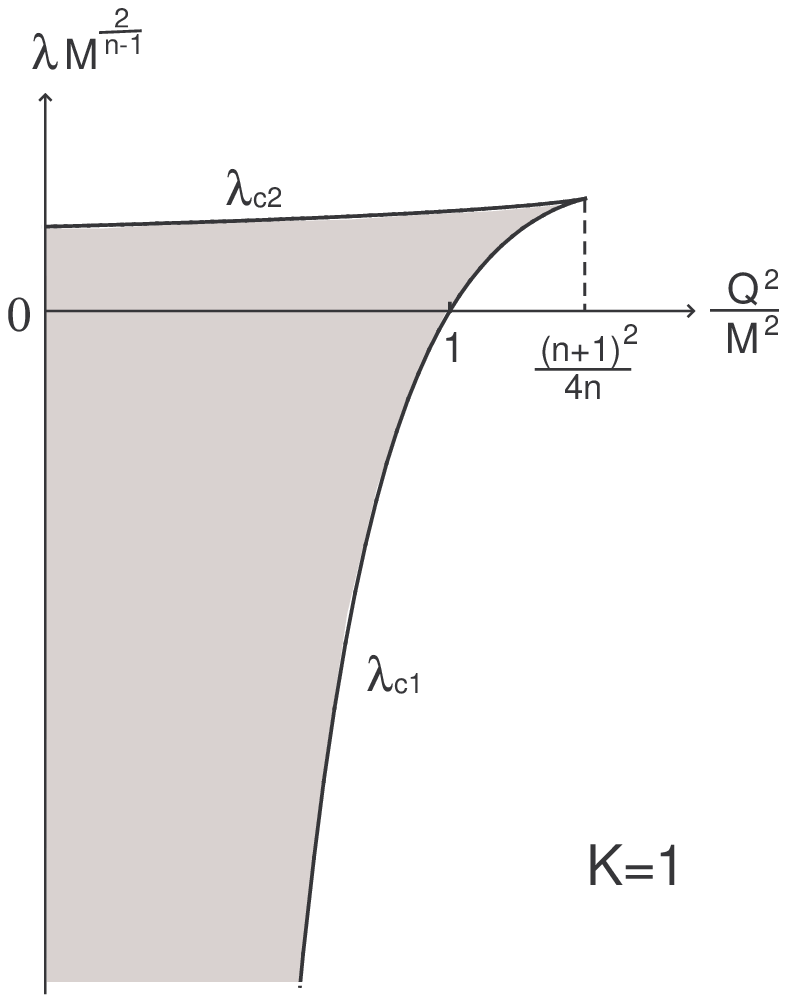}}
\end{minipage}
\begin{minipage}{4cm}
\centerline{\includegraphics[width=4cm]{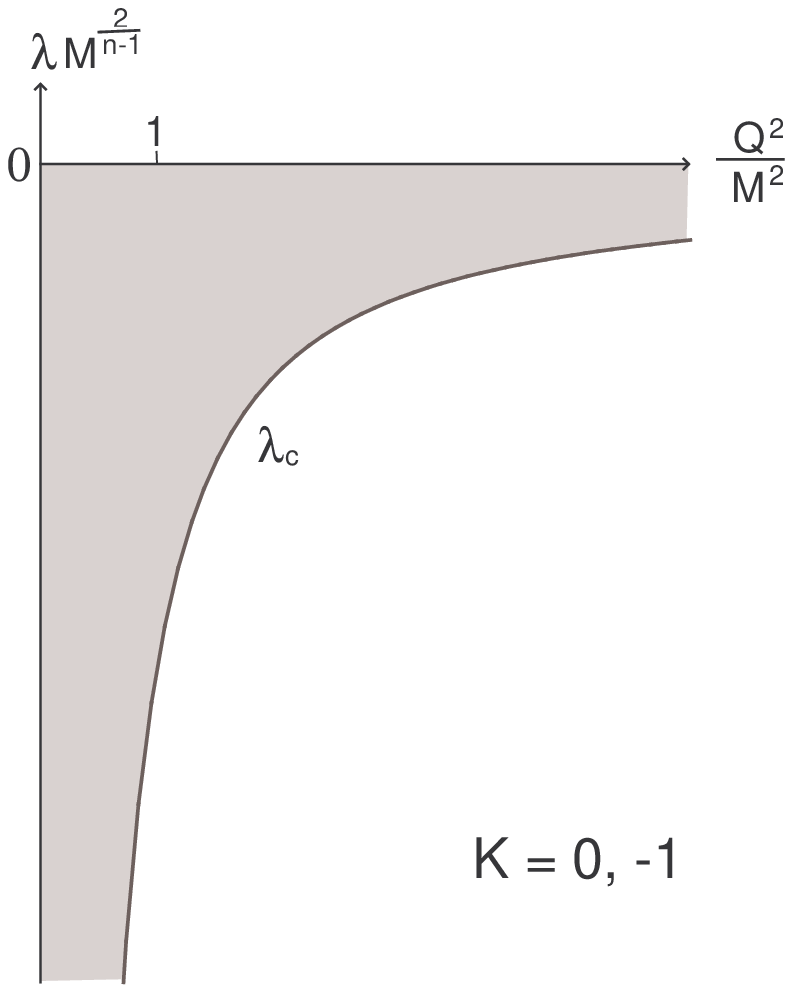}}
\end{minipage}
\caption{The parameter ranges in which the spacetime contains a 
regular black hole (the shaded regions). }\label{fig:BHcondition}
\end{figure}

In general, this solution have a naked singularity or does not have 
a horizon. As shown in Ref.\cite{Kodama.H&Ishibashi2004}, it gives a 
regular black hole spacetime only when the parameters $M,Q$ and 
$\lambda$ are in the special regions depicted in 
Fig.\ref{fig:BHcondition}. In this case, the Einstein space $\K^n$ 
describes a spatial section of the black hole horizon and at the same 
time the spatial infinity.

\subsection{Uniqueness of black holes}

In the static case with $\Lambda=0$, the uniqueness of 
asymptotically flat regular black holes is now established in higher 
dimensions as well: all regular solutions are exhausted by the 
Tangherlini-Schwarzschild solution for the vacuum 
system\cite{Hwang.S1998,Gibbons.G&Ida&Shiromizu2003} and by the 
higher-dimensional Reissner-Norsdtr\"om solution given by 
\eqref{GeneralisedStaticSol} and 
\eqref{GeneralisedStaticSol:f&E} with $K=1$ and $\Lambda=0$ in the 
non-degenerate case\cite{Gibbons.G&Ida&Shiromizu2002a} and the 
higher-dimensional Majumdar-Papapetrou solutions in the degenerate 
case for the Einstein-Maxwell system\cite{Rogatko.M2003}. Further, 
it is also proved that the Gibbons-Maeda solution and the 
Tangherlini-Schwarzschild solution are the only asymptotically flat 
regular black hole solutions for the Einstein-Maxwell-Dilaton system 
and for the Einstein-Harmonic-Scalar system, 
respectively\cite{Gibbons.G&Ida&Shiromizu2002b,Rogatko.M2002}. 
However, for $\Lambda\not=0$, nothing is known about the uniqueness, 
although it was pointed out in Ref. 
\cite{Anderson.M&Chrusciel&Delay2002} that the approach developed by 
Anderson may also apply to higher-dimensional cases with 
$\Lambda<0$. 

In the rotating case, the higher-dimensional counter-part of the 
Kerr solution is also known. It is given by the Myers-Perry 
solution\cite{Myers.R&Perry1986}, which has a spherical horizon and 
is asymptotically flat. In contrast to the static case, however, 
this is not the unique asymptotically flat rotating regular 
solutions in higher dimensions, although the uniqueness holds for 
supersymmetric black holes in the 5-dimensional $N=1$ minimal 
supergravity model\cite{Reall.H2003}. This is because the horizon 
topology need not be given by the sphere in higher-dimensions as 
discussed in Ref. \cite{Cai.M&Galloway2001}. In fact, Emparan and 
Reall\cite{Emparan.R&Reall2002a} found an asymptotically flat and 
rotating regular black hole solution in five dimensions whose black 
hole surface is homeomorphic to $S^2\times S^1$. Thus, at present, we 
have two different families of asymptotically flat rotating regular 
solutions with different horizon topologies. Since more complicated 
horizon topology is allowed in dimensions greater than 5, it is 
highly probable that other families of regular solutions exist. 
However, near the static and spherically symmetric limit, 
it is likely that a uniqueness theorem holds, as we will see later. 

\subsection{Stability of black holes}

Recently, the author and Ishibashi have shown that for a static 
charge black hole in higher dimensions represented by 
\eqref{GeneralisedStaticSol} and \eqref{GeneralisedStaticSol:f&E}, 
perturbation equations can be also reduced to decoupled 2nd-order 
master equations of the Schr\"odinger type, as in four 
dimensions\cite{Kodama.H&Ishibashi2003,Kodama.H&Ishibashi2004}. 
These master equations in higher dimensions, however, have some new 
features. First, the effective potential $V$ is not positive 
definite in general. Hence, there may exist an unstable mode. 
Second, there exists no simple relation between vector and scalar 
perturbations like the scalar-vector correspondence in four 
dimensions, for $d=n+2>4$. This implies that stabilities for scalar 
and vector perturbations should be studied separately. Third, there 
exist tensor perturbations for $d=n+2>4$.

Now, let us see how these new features affect the stability of 
static black holes in higher dimensions.

\subsubsection{Tensor perturbations}

When $\K^n$ is a generic Einstein space, the $\K^n$-coordinates 
appear in  perturbation equations only through the Lichnerowicz 
operator
\Eq{
\hat \triangle_L h_{ij}=-\hat D\cdot\hat D h_{ij}-2\hat R_{ikjl}h^{kl}
   +2(n-1)K h_{ij}.
}
Hence, by expanding tensor perturbations in terms of the 
eigentensors of $\hat \triangle_L$,
\Eq{
\hat \triangle_L \THB_{ij}=\lambda_L \THB_{ij};\quad
\THB^i_i=0,\quad \hat D^j\THB_{ij}=0,
}
we obtain a single decoupled equation for $H_T^{(2)}$, which can be 
easily put into the form \eqref{MasterEq}. 

In the special case in which $\K^n$ is a constant curvature space, 
$\hat\triangle_L$ is expressed as $\hat\triangle_L=-\hat D\cdot\hat 
D+2nK$. Hence, the expansion by $\THB_{ij}$ is identical to harmonic 
expansion, and the eigenvalue $\lambda_T$ is related to the 
eigenvalue $k_T^2$ with respect to $-\hat D\cdot\hat D$ by 
$\lambda_L=k_T^2+2nK$. Note that $k_T^2$ is always non-negative and 
takes the discrete value $k_T^2=l(l+n-1)-2\; (l=2,3,\cdots)$ for 
$\K^n=S^n$ ($K=1$).

For neutral black holes, the effective potential for tensor perturbations reads
\Eq{
 V_T = \frac{f}{r^2}  
   \left[\lambda_L-(3n-2)K+
      \frac{n(n+2)}{4}f+\frac{n(n+1)M}{r^{n-1}}
         \right].  
}
When $\K^n$ is maximally symmetric, $V_T>0$ outside the horizon 
since $\lambda_L-(3n-2)K=k_T^2-(n-2)K\ge0$. Hence, the black hole is 
stable for a tensor perturbation. In contrast, when $\K^n$ is a 
generic Einstein space, no general lower bound for $\lambda_L$ is 
known, and generalised static black holes can be unstable against a 
tensor perturbation\cite{Gibbons.G&Hartnoll2002}.

Next for charged black holes, since  EM field perturbations are not 
coupled to  tensor perturbations of the metric, the master equation 
is again given by a single equation. The effective potential is 
given by
\Eqr{
&V_T=
 &\frac{f}{r^2}\Big[ \lambda_L-(3n-2)K+\frac{n(n+2)f}{4}
   +\frac{n(n+1)M}{r^{n-1}} \notag\\
&&\qquad\quad -\frac{n^2Q^2}{r^{2n-2}} \Big].
}
We can show that this potential is positive definite outside the 
horizon if $\lambda_L-2(n-1)K=k_T^2+2K\ge0$. Hence, when $\K^n$ is 
maximally symmetric, the black hole is stable against tensor 
perturbations for $K\ge0$. However, for $K=-1$ ($\lambda<0$), the 
potential becomes negative near the horizon if $\lambda$ is close to 
the lower limit $\lambda_{c-}$ in Fig.\ref{fig:BHcondition}. Hence, 
the black hole might be unstable in this case even when $\K^n$ is 
maximally symmetric.

\subsubsection{S-deformation}

For scalar and vector perturbations, we can expand perturbation 
variables by scalar harmonics $\SHB$ and vector harmonics $\VHB_i$ 
on $\K^n$ satisfying $\hat \triangle \SHB =-k^2 \SHB$ and $\hat 
D\cdot\hat D \VHB_i=-k_V^2\VHB_i$. For each harmonic,  there appear 
two modes of perturbations, an electromagnetic mode $\Phi_+$ and a 
gravitational mode $\Phi_-$. They obey two decoupled equations of 
the form \eqref{MasterEq:RN}, as in the case of 4D 
Reissner-Nordstr\"om black hole. However, $V_\pm$ is not positive 
definite when $n$ is large, even for spherically symmetric black 
hole. In order to study the stability for such a system, we utilised 
the following method that we call {\em the $S$-deformation method}.

Let $I$ be the range of $r_*$ corresponding to the regular region 
outside of the horizon, $-\infty <r_*<r_{*\infty}$. Here, 
$r_{*\infty}$ is $+\infty$ for $\lambda\ge0$, but is finite for 
$\lambda<0$.  Then, in the space $C_0^\infty(I)$ of smooth functions 
with compact support, the operator
\Eq{
A:=-\frac{d^2}{dr_*^2} + V(r)
}
is symmetric. We assume that $A$ is extended to a self-adjoint 
operator in  $L^2(I)$ by the Friedrichs extension. Then, the lower 
bound for the spectrum of $A$ coincides with the lower bound of 
$\omega^2=(\Phi,A\Phi)/(\Phi,\Phi)$ ($\Phi\in C^\infty_0(I)$). Here, 
for any regular function $S(r)$ on $I$, we have 
\Eqr{
(\Phi,A\Phi)
  =\int_I dr_*\left[ 
  \left|\left(\frac{d}{dr_*}+S(r)\right)\Phi\right|^2 
             + \tilde V|\Phi|^2\right],
}
where 
\Eq{
\tilde V=V + f\frac{dS}{dr} -S^2.
}
Hence, if we can show that $\tilde V$ is non-negative for an 
appropriate $S$, we can conclude that $\omega^2\ge0$, i.e., the 
stability of the system.

\begin{figure}
\begin{minipage}{4cm}
\includegraphics[height=5cm]{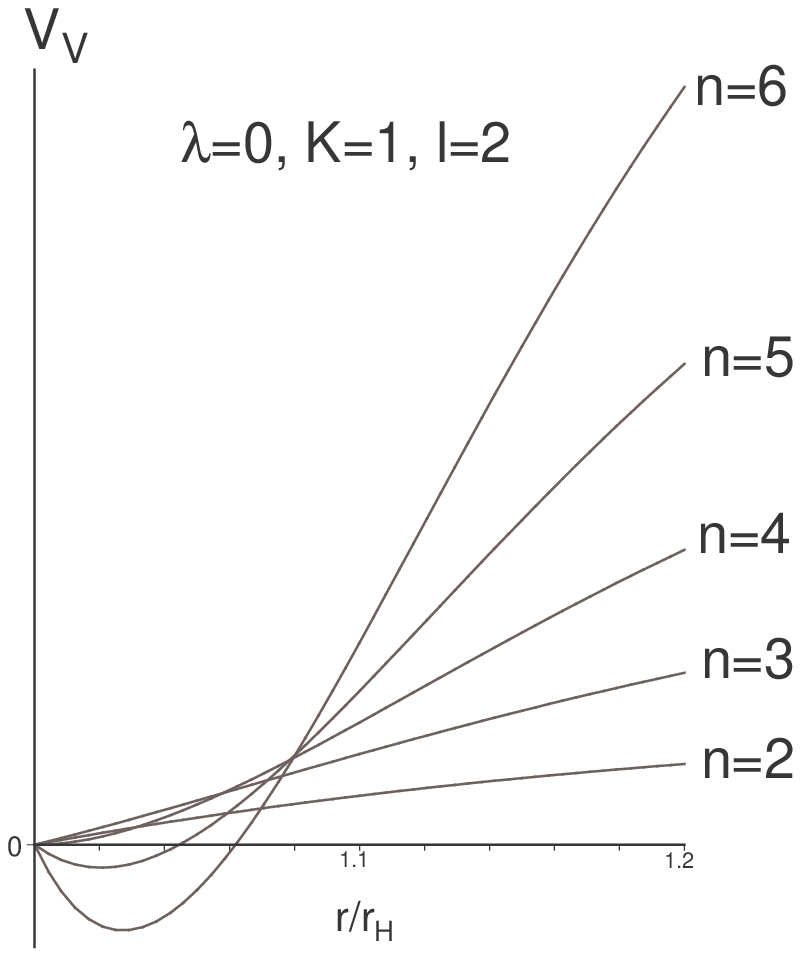}
\end{minipage}
\begin{minipage}{4cm}
\includegraphics[height=5cm]{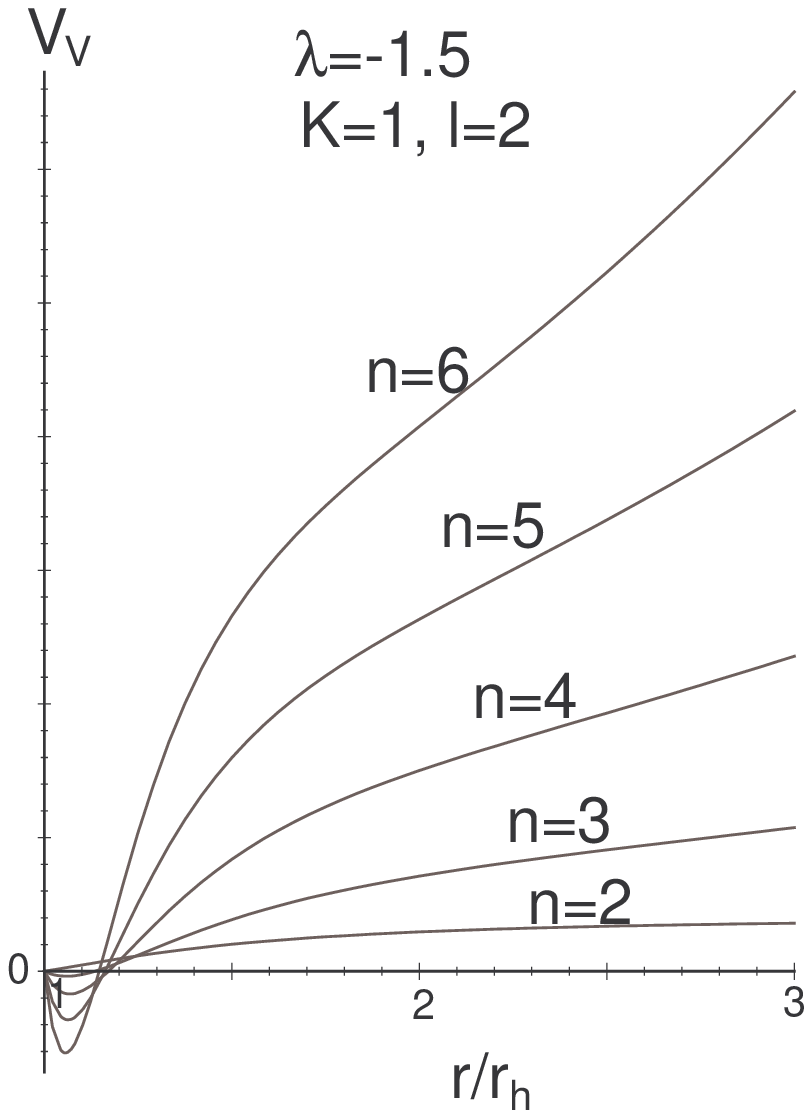}
\end{minipage}
\caption{Examples of $V_V=V_{V-}$ for $Q=0$.}\label{fig:V_V}
\end{figure}

\subsubsection{Vector perturbations}

The effective potentials for vector perturbations are given by
\Eqr{
&V_{V\pm}
 =&\frac{f}{r^2}\Big[k_V^2 +\frac{(n^2-2n+4)K}{4}
   -\frac{n(n-2)}{4}\lambda r^2 \notag\\
&& +\frac{n(5n-2)Q^2}{4r^{2n-2}} 
   +\frac{-(n^2+2)M\pm 2\Delta}{2r^{n-1}}\Big],
}
where
\Eq{
\Delta^2= (n^2-1)^2M^2+2n(n-1)[k_V^2-(n-1)K] Q^2.
}
These potentials are not positive definite when $n$ is large as 
illustrated in Fig.\ref{fig:V_V}.

Nevertheless, we can show that there exists no unstable mode for 
most cases, by using the $S$-deformation method. In fact, for $ 
S=nf/2r$, the $S$-deformation yields
\Eq{
\tilde V_{V\pm} =\frac{f}{r^2}\left[ m_V
   +\frac{(n^2-1)M\pm\Delta}{r^{n-1}} \right],
}
where $m_V=k_V^2-(n-1)K$. Since $m_V\ge0$, $\tilde V_+$ is positive 
definite. Hence, the EM mode of a vector perturbation is always 
stable. Further, for $K\ge0$, we can also show that $\tilde V_-$ is 
positive definite. Hence, we can conclude that black holes with 
$K\ge0$ are stable against vector perturbations. In contrast, for 
$K=-1$ ($\lambda<0$), $\tilde V_-$ becomes negative near the horizon 
when $\lambda$ is close to the critical value $\lambda_{c-}$. In 
this case, we cannot prove the stability by this method.

\begin{figure}
\begin{minipage}{4cm}
\includegraphics[width=4cm]{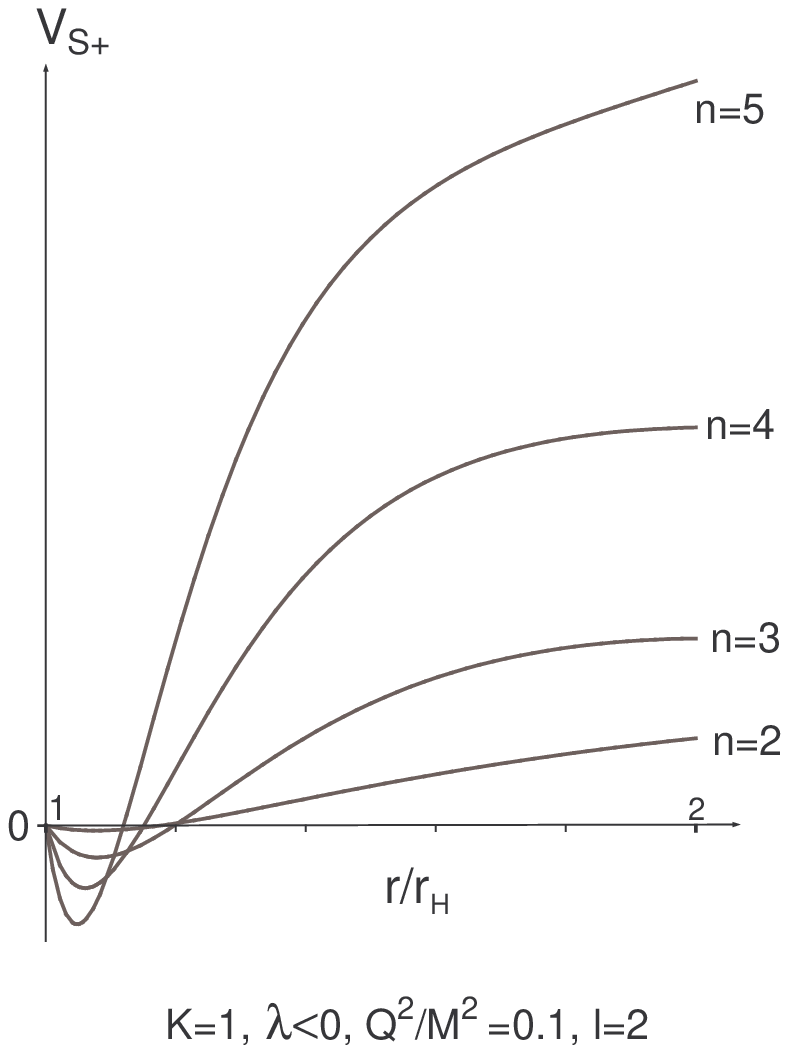}
\end{minipage}
\begin{minipage}{4cm}
\includegraphics[width=5cm]{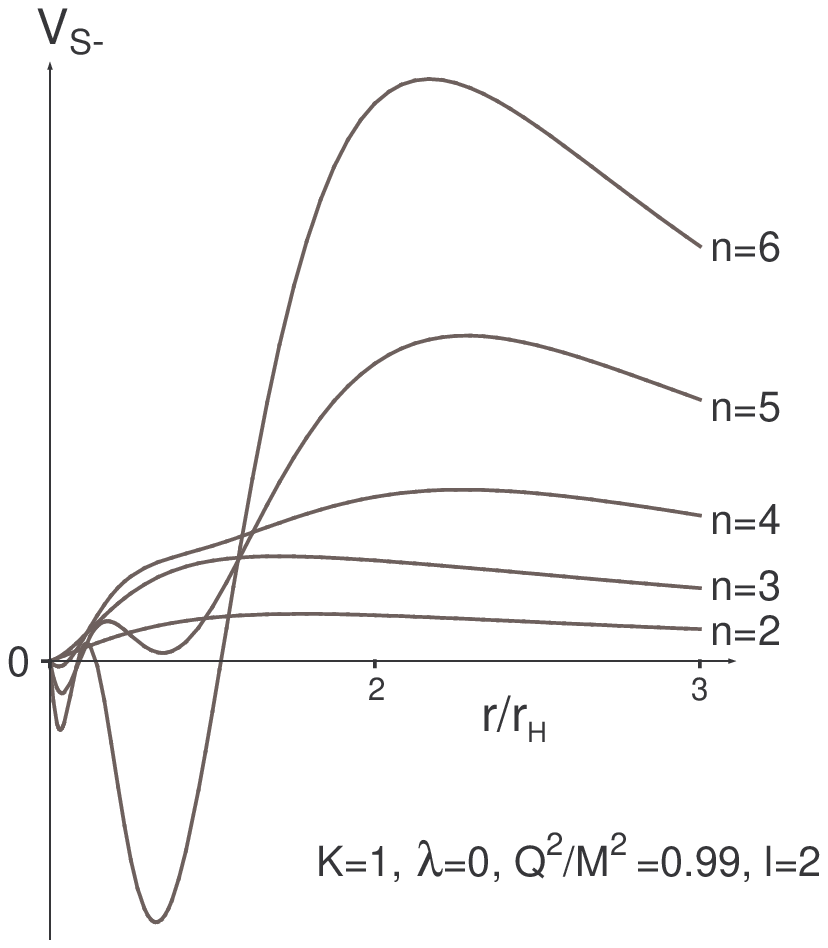}
\end{minipage}
\caption{Examples of $V_{S\pm}$}\label{fig:V_S}
\end{figure}

\subsubsection{Scalar perturbations}

The explicit expressions of the effective potentials $V_{S\pm}$ for 
scalar perturbations are quite long. So, we only give the graphs of 
them for some special parameter values in Fig. \ref{fig:V_S}. As 
these figures show, either of $V_{S+}$ and $V_{S-}$ is not positive 
definite in general.

However, by the $S$-deformation method again, we can show that the 
black hole is always stable for the EM mode 
perturbation\cite{Kodama.H&Ishibashi2004}. We can also show that 
asymptotically flat neutral black holes in arbitrary 
dimensions\cite{Ishibashi.A&Kodama2003} and asymptotically dS and 
AdS neutral black holes in four dimensions are 
stable\cite{Kodama.H&Ishibashi2003}. However, the stability of 
charged black holes has been proved only for special cases. The 
present status of the perturbative analysis of generalised static black 
holes is summarised in Table \ref{tbl:stability}.

{\squeezetable
\begin{table}[t]
\begin{tabular}{l|l|c|c|c|c|c|c|}
\hline\hline
\multicolumn{2}{c|}{}& \multicolumn{2}{c|}{Tensor}
  & \multicolumn{2}{c|}{Vector}& 
\multicolumn{2}{c|}{Scalar}\\\cline{3-8}
\multicolumn{2}{c|}{}&$Q=0$ & $Q\not=0$ &$Q=0$ & $Q\not=0$ 
&$Q=0$ & $Q\not=0$ \\
\hline
$K=1$& $\lambda=0$ & OK & OK & OK & OK 
     & OK 
     & $\begin{array}{l} 
         d=4,5\ \text{OK} \\ d\ge6\ \text{?} 
       \end{array}$
     \\
\cline{2-8}
     &$\lambda>0$ & OK & OK & OK & OK 
     & $\begin{array}{l} 
         d\le6\ \text{OK} \\ d\ge7\ \text{?} 
        \end{array}$
     & $\begin{array}{l}
         d=4,5\ \text{OK} \\ d\ge6\ \text{?} 
        \end{array}$
     \\
\cline{2-8}
     &$\lambda<0$ & OK & OK & OK & OK 
     &  $\begin{array}{l}
          d=4\ \text{OK} \\ d\ge5\ \text{?} 
         \end{array}$
     &  $\begin{array}{l}
          d=4\ \text{OK} \\ d\ge5\ \text{?} 
         \end{array}$
     \\
\hline
$K=0$ &$\lambda<0$ & OK & OK & OK & OK 
     & $\begin{array}{l}
         d=4\ \text{OK} \\ d\ge5\ \text{?} 
        \end{array}$
     & $\begin{array}{l}
         d=4\ \text{OK} \\ d\ge5\ \text{?} 
       \end{array}$ 
     \\
\hline
$K=-1$ &$\lambda<0$ & OK & ? & OK & ? 
     & $\begin{array}{l}
         d=4\ \text{OK} \\ d\ge5\ \text{?} 
        \end{array}$
     & $\begin{array}{l}
         d=4\ \text{OK} \\ d\ge5\ \text{?} 
        \end{array}$
     \\
\hline
\end{tabular}
\caption{Stability of generalised static black holes}
\label{tbl:stability}
\end{table}
}

\subsection{Perturbative uniqueness}

As explained in \S\ref{PerturbativeUniqueness:4D}, we can study the 
perturbative uniqueness of black holes near the static limit by 
looking for bounded stationary solutions to the perturbation 
equations. Actually, we can establish a kind of perturbative 
uniqueness theorem by this approach. For the limitation of space, we 
only give an outline of the argument here. The detail will be given 
in another paper\cite{Kodama.H2004}.

First of all, note that there exist such solutions with $l=1$ for 
the vector perturbation equation in the spherically symmetric case 
($K=1$) as shown in Ref.\cite{Ishibashi.A&Kodama2003}. To be 
precise, vector harmonics on $S^n$ are in one-to-one correspondence 
to Killing vectors of $S^n$, and for each harmonic there exists one 
stationary solution to the vector perturbation equation. A Killing 
vector on $S^n$ is further in one-to-one correspondence to an 
antisymmetric matrix of rank $n+1$, and the conjugate class of this 
matrix is classified by its $N$ eigenvalues, where $N=[(n+1)/2]$ is 
the rank of $\SO(n+1)$. Hence, these stationary solutions exactly 
correspond to the angular momentum freedom of the Myers-Perry 
solution\cite{Myers.R&Perry1986}.

In the asymptotically flat case, we can show that there exists no 
other stationary solution for which $\delta g_{\mu\nu}$ are bounded 
everywhere\cite{Kodama.H&Ishibashi2003}. Hence, the Myers-Perry 
solution is the unique regular solution near the static limit. In 
the asymptotically de Sitter case, we can obtain the same result for 
the tensor and vector perturbations by using the S-deformation 
technique. The perturbative uniqueness for the scalar perturbation 
can also be proved in arbitrary dimensions by applying a similar 
argument to a master equation that holds only for stationary 
perturbations. In contrast, the asymptotically anti-de Sitter case 
is more subtle. In this case, there exist infinitely many solutions 
with bounded $\delta g_{\mu\nu}$ in addition to the $l=1$ vector 
perturbations, in accordance with the general result in the four 
dimensional case\cite{Anderson.M&Chrusciel&Delay2002}. However, if 
we require the stronger but natural asymptotic condition that every 
component of the metric perturbation with respect to an background 
orthonormal basis  vanishes at $r\sim\infty$, these extra solutions 
are excluded. 

\section{Concluding Remarks}

In this article, we have overviewed the present status of the 
uniqueness and stability issue of black holes in four and higher 
dimensions. We have seen that the uniqueness and stability are well 
established for asymptotically flat neutral and static black holes 
in arbitrary dimensions and for asymptotically flat stationary black 
holes in four dimensions, but the situation concerning the other 
cases is quite unsatisfactory. In particular, the classification and 
stability analysis of rotating black holes and asymptotically de 
Sitter or anti-de Sitter black holes in higher dimensions are 
important open problems, although the perturbative analysis can 
give some useful information as we have shown. In order to solve 
these problems in the nonperturbative framework, it will be 
necessary to find a higher-dimensional extension of the rigidity 
theorem.

\section*{Acknowledgement}

This work is partly supported by the JSPS grant No. 
15540267.


\end{document}